# Reduction in nuclear size and quadrupole deformation of high-spin isomers of 127,129In

A.R. Vernon,[1, 2, 3, *] C.L. Binnersley,[1] R.F. Garcia Ruiz ,[2, 4, †] K.M. Lynch ,[1, 5] T. Miyagi ,[6, 7, 8, 9, 10] J. Billowes,[1] M.L. Bissell,[1] T.E. Cocolios ,[3] J.P. Delaroche,[11, 12] J. Dobaczewski ,[13, 14] M. Dupuis,[11, 12] K.T. Flanagan,[1] W. Gins ,[3] M. Girod,[11] G. Georgiev,[15] R.P.de Groote ,[3, 16] J. D. Holt ,[10, 17] J. Hustings,[3] Á. Koszorús ,[3] D. Leimbach ,[4] J. Libert,[11] W. Nazarewicz ,[18, 19] G. Neyens,[3] N. Pillet,[11] P.-G. Reinhard ,[20] S. Rothe,[4] B.K. Sahoo ,[21] S. R. Stroberg ,[22] S.G. Wilkins ,[2, 4] X.F. Yang ,[3, 23] Z.Y. Xu,[3] and D.T. Yordanov [24]

[1] *Department of Physics and Astronomy, The University of Manchester, Manchester M13 9PL, United Kingdom*
[2] *Massachusetts Institute of Technology, Cambridge, MA 02139, USA*
[3] *KU Leuven, Instituut voor Kern- en Stralingsfysica, B-3001 Leuven, Belgium*
[4] *CERN, CH-1211 Geneva 23, Switzerland*
[5] *EP Department, CERN, CH-1211 Geneva 23, Switzerland*
[6] *Center for Computational Sciences, University of Tsukuba, Ibaraki, 305-8577, Japan*
[7] *Technische Universität Darmstadt, Department of Physics, 64289 Darmstadt, Germany*
[8] *ExtreMe Matter Institute EMMI, GSI Helmholtzzentrum für Schwerionenforschung GmbH, 64291 Darmstadt, Germany*
[9] *Max-Planck-Institut für Kernphysik, Saupfercheckweg 1, 69117 Heidelberg, Germany*
[10] *TRIUMF, Vancouver, BC V6T 2A3, Canada*
[11] *CEA, DAM, DIF, F-91297 Arpajon, France*
[12] *Université Paris-Saclay, CEA, LMCE, 91680 Bruyères-le-Châtel, France*
[13] *School of Physics, Engineering and Technology,
University of York, Heslington, York YO10 5DD, United Kingdom*
[14] *Institute of Theoretical Physics, Faculty of Physics,
University of Warsaw, ul. Pasteura 5, PL-02-093 Warsaw, Poland*
[15] *IJCLab, CNRS-IN2P3, Université Paris-Sud, Université Paris-Saclay, Orsay, France*
[16] *Department of Physics, University of Jyväskylä, Survontie 9, Jyväskylä, FI-40014, Finland*
[17] *Department of Physics, McGill University, Montréal, QC H3A 2T8, Canada*
[18] *Facility for Rare Isotope Beams, Michigan State University, East Lansing, Michigan 48824, USA*
[19] *Department of Physics and Astronomy, Michigan State University, East Lansing, Michigan 48824, USA*
[20] *Institut für Theoretische Physik II, Universität Erlangen-Nürnberg, 91058 Erlangen, Germany*
[21] *Atomic, Molecular and Optical Physics Division,
Physical Research Laboratory, Navrangpura, Ahmedabad 380009, India*
[22] *Department of Physics and Astronomy, University of Notre Dame, Notre Dame, IN 46556, USA*
[23] *School of Physics and State Key Laboratory of Nuclear Physics and Technology, Peking University, Beijing 100871, China*
[24] *IJCLab, Université Paris-Sud, Université Paris-Saclay, Orsay, France*

We employed laser spectroscopy of atomic transitions to measure the nuclear charge radii and electromagnetic properties of the high-spin isomeric states in neutron-rich indium isotopes ($Z = 49$) near the closed proton and neutron shells at $Z = 50$ and $N = 82$. Our data reveal a reduction in the nuclear charge radius and intrinsic quadrupole moment when protons and neutrons are fully aligned in $^{129}$In($N = 80$), to form the high spin isomer. Such a reduction is not observed in $^{127}$In($N = 78$), where more complex configurations can be formed by the existence of four neutron-holes. These observations are not consistently described by nuclear theory.

*Introduction.* Excited states of atomic nuclei most commonly occur with extremely short lifetimes, typically ranging from less than a picosecond ($10^{-12}$ s) to a few nanoseconds ($10^{-9}$ s)[1]. However, in some nuclei, the protons and neutrons can be reorganized to form configurations with exceptionally large differences in shape or angular momentum. These states can increase in their lifetime due to quantum selection rules that greatly inhibit their de-excitation pathways [1–4], giving rise to long-lived states - nuclear isomers. Isomers formed by three or more unpaired nucleons can create nuclear states with unusually large values of angular momentum (high nuclear spin, $I$), commonly known as multi-quasiparticle isomers [1]. These isomers can exhibit lifetimes that are comparable to or even longer than those of their respective ground states (g.s.) [5–9]. Until now, there are only a handful of high-spin, multi-quasiparticle isomers for which properties such as the charge radii and electromagnetic moments have been measured simultaneously. Available data exists only for some heavy nuclei, e.g. $^{97}$Y, $^{130}$Cs, $^{130}$Ba, $^{177}$Lu, and $^{178}$Hf [10–13]. All of these isomers were found to have smaller charge radii than their g.s. while exhibiting a similar or larger quadrupole moment. Accurate calculations of these isomers are currently beyond the reach of many microscopic calculations, as they exhibit complex configurations [13–17]. This calls for measurements of high-spin isomers in the vicinity of closed-shell nuclei that are expected to have simpler configurations accessible to complementary nuclear models [13, 18]. In these systems, pairing correlations may play a lesser role, and thus, the standard hypothesis that links the isomer shift to the pairing-blocking effect [1] can be critically evaluated and tested. Such isomers exist in the neutron-rich indium isotopes ($Z = 49$), near the doubly-magic $^{132}$Sn ($Z = 50$, $N = 82$), which recently became accessible for high-precision laser spectroscopy studies [19]. These recent measurements of the ground-state properties, including charge radii, for the indium isotopes have established the credibility of the currently available theoretical calculations [19, 20].

In this work, we report measurements of the changes in mean-squared charge radii and electromagnetic moments of the high-spin isomers (nuclear spin suggested to be $I \geq 21/2$) of neutron-rich indium isotopes. In the shell-model picture, the g.s. of the odd-even indium isotopes is dominated by a hole in the $Z = 50$ proton closed shell ($g_{9/2}$ proton orbit), resulting in a nuclear spin $I^\pi = 9/2^+$. This is the case for all known odd-even isotopes[19]. However, beta-decay spectroscopy studies revealed the existence of high-spin, multi-quasiparticle isomers, $I^\pi = (21/2)^-$ and $I^\pi = (23/2)^-$, in $^{127}$In ($N = 78$) and $^{129}$In ($N = 80$), respectively [21, 22]. These isomers are suggested to be formed by the breaking of a neutron pair to create single holes in the $d_{3/2}$ and $h_{11/2}$ neutron orbits, interacting with a proton hole in the $g_{9/2}$ orbit [21]. A schematic diagram of the proton and neutron configuration of these isomers is shown in Fig. 1(a). The nuclear spin 23/2 is obtained by fully aligning the spins of the 3 hole states in the $^{132}$Sn core, adding up to $I = 23/2$. This maximally aligned, or optimal, structure of the isomer in $^{129}$In, presents a simple configuration to guide our understanding of these nuclei. In $^{127}$In, a different spin assignment, 21/2, has been suggested in the literature [21, 22], which corresponds to a configuration where not all of the 3 unpaired holes are fully aligned. Therefore, the measured spectra have been analyzed assuming the two possible spin assignments for $^{127}$In, 21/2 and 23/2.

*Experimental results.* Our measurements of the high-spin isomers were enabled by recent developments in the sensitive method of collinear laser spectroscopy at radioactive ion beam facilities (see Supplemental Material), which allowed high-resolution measurements of the high-spin isomers produced in a high background of other long-lived species, such as other isotopes or molecules with similar mass. The neutron-rich $^{127,129}$In isotopes and their high-spin isomeric states were produced at the Isotope Separation Online facility (ISOLDE) at CERN and directed to the Collinear Resonance Ionization Spectroscopy (CRIS) experiment [23] for sensitive measurements of their hyperfine structure. The hyperfine spectra of $^{127,129}$In were measured in the 5p $^2P_{1/2} \to$ 8s $^2S_{1/2}$ (246.0 nm) and 5p $^2P_{3/2} \to$ 9s $^2S_{1/2}$ (246.8 nm) transitions of the indium atom. A major experimental challenge in laser spectroscopy studies of high-spin isotopes is the population of several possible hyperfine states, which causes a highly congested spectrum. Thus, we use the transition from the low electronic state, $^2P_{1/2}$, which does not exhibit quadrupole hyperfine splitting, to obtain the nuclear magnetic moment. Subsequently, once the magnetic moment is fixed, we use a transition from the state $^2P_{3/2}$ to extract the nuclear quadrupole moment. More details of the experimental setup alongside examples of the measured spectra can be found in the Supplemental Material.

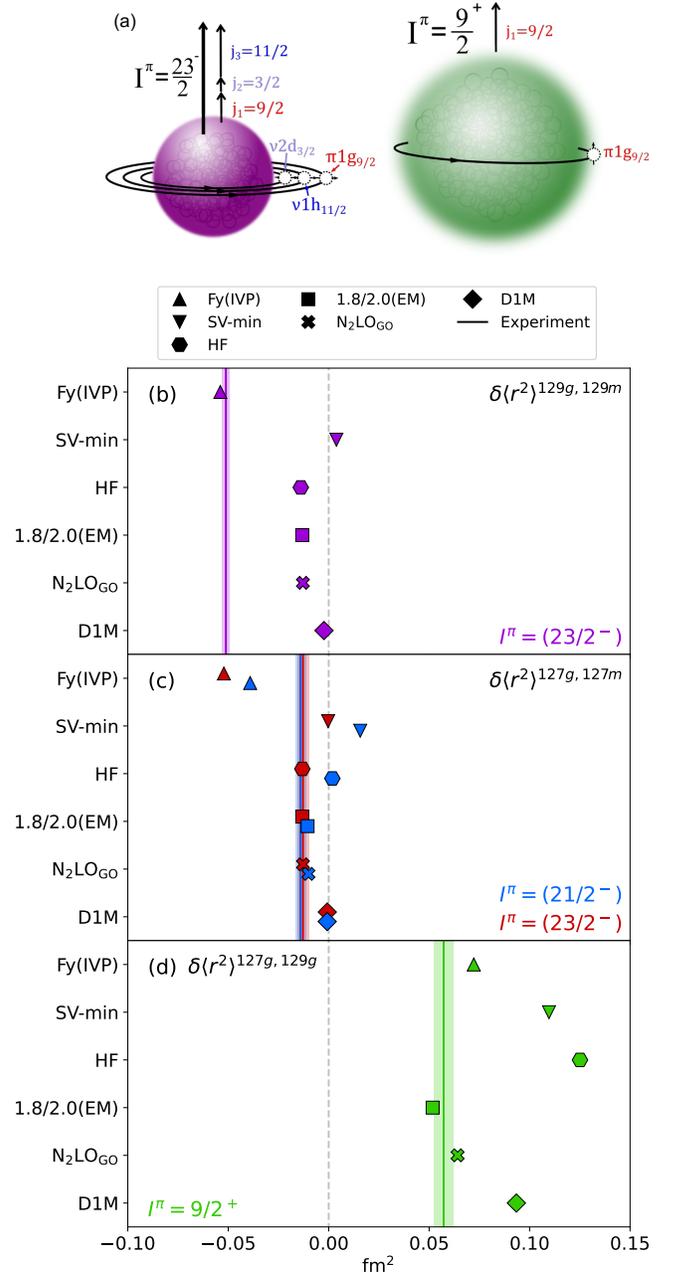

FIG. 1. State-dependent charge radii of the neutron-rich indium isotopes. (a) Shell-model configurations associated with the $I^\pi = 9/2^+$ g.s. and $I^\pi = (23/2^-)$ high-spin isomer in $^{129}$In. (b-d) Change in charge radii of the ground state and high-spin states in indium isotopes. Experimental data (lines) are compared with results of the DFT (Fy(IVP), SV-min, and HF), VS-IMSRG (1.8/2.0(EM) and N$_2$LO$_{GO}$) and MPMH (D1M) calculations (markers). (b) $\delta \langle r^2 \rangle^{g,m}$ of the high-spin (23/2$^-$) isomer of $^{129}$In relative to the 9/2$^+$ g.s. (c) $\delta \langle r^2 \rangle^{g,m}$ of the high-spin isomer of $^{127}$In relative to the 9/2$^+$ g.s. assuming $I^\pi = (21/2^-)$ (blue) and (23/2$^-$) (red) for the isomer. (d) Change in g.s. charge radii $\delta \langle r^2 \rangle^{127,129}$ from Ref. [20]. The vertical lines show the experimental values corresponding to the weighted average of the $\delta \langle r^2 \rangle$ values for the 5p $^2P_{1/2} \to$ 8s $^2S_{1/2}$ and 5p $^2P_{3/2} \to$ 9s $^2S_{1/2}$ transitions, with the uncertainty given by the shaded area.

The isomer shift, $\delta\nu^{g,m}$, measured as the difference between the centroid frequency of the hyperfine structure of the ground- and isomeric state, is related to the change of their root-mean-squared charge radii, $\delta\langle r^2\rangle^{g,m}$, using the relation

$$\delta\nu^{g,m} = F\delta\langle r^2\rangle^{g,m} + K^{MS}\frac{m_m - m_g}{m_g m_m}, \qquad (1)$$

where $m_g$ and $m_m$ are the masses of the ground and high-spin states of the $^{127,129}$In isotopes [24, 25], respectively, $F$ and $K^{MS} = K^{SMS} + K^{NMS}$ are the field-shift (FS) and mass-shift (MS) constants, obtained from recently-developed atomic calculations, as discussed in Refs. [20, 26]. Similarly, the isotope shift between $^{129}$In and $^{127}$In were obtained from the frequency shift between isotopes, $\delta\nu^{127,129}$. The differential charge radii in Eq. (1) are defined as: $\delta\langle r^2\rangle^{A,A'} \equiv \langle r_c^2\rangle^{A'} - \langle r_c^2\rangle^A$.

The $\delta\nu^{g,m}$ and $\delta\nu^{127,129}$ measurements and extracted $\delta\langle r^2\rangle^{g,m}$ and $\delta\langle r^2\rangle^{127,129}$ values are displayed in Table I. The average values presented are a weighted average of the mean squared charge radii extracted for the two atomic transitions, weighted by their statistical uncertainties. The largest systematic uncertainty (of the two atomic transitions) is quoted alongside. The ground and high-spin states could be identified by their relative intensity and electromagnetic moments. The spin of the high-spin isomers, $(21/2)^-$ for $^{127}$In and $23/2^-$ for $^{129}$In, were taken from $\beta\gamma$-coincidence measurements and supported by theoretical predictions [21, 27–29]. We present results obtained by assuming both $(21/2)^-$ and $(23/2)^-$ spins for $^{127}$In.

The spectroscopic nuclear electric quadrupole moments, $Q_S$, were extracted from the measured hyperfine quadrupole constants, $B_{hf}$, using the relation

$$B_{hf} = e\, Q_S\, V_{zz}, \qquad (2)$$

where a value of $B_{hf}(^2P_{3/2})/Q_S = 576(4)$ MHz/b obtained from relativistic coupled-cluster calculations was used [19, 31]. The magnetic moments, $\mu$, were determined from the magnetic hyperfine constant, $A_{hf}$, using a reference NMR value of $\mu_{ref} = +5.5408(2)$ $\mu_N$[32], and the measured $A_{ref}$ for the isotope $^{115}$In [19, 33], according to

$$\mu = \mu_{ref}\frac{I A_{hf}}{I_{ref}, A_{ref}}(1 + \Delta), \qquad (3)$$

where the differential hyperfine anomaly, $\Delta$, is not considered as it is expected to be smaller than our experimental uncertainty for these isotopes [34].

*Theoretical results.* We compare our experimental $\delta\langle r^2\rangle$ values to nuclear structure calculations, performed using three complementary theoretical methods: (i) valence-space in-medium similarity renormalization Group (VS-IMSRG) method [35–37]; (ii) Multi-Configuration Self-Consistent Field approach (MPMH) [38–42]; and (iii) Density Functional Theory (DFT) [19, 43, 44]. The extracted $\delta\langle r^2\rangle^{g,m}$ and $\delta\langle r^2\rangle^{127,129}$ values are shown alongside calculated values in Fig. 1.

The VS-IMSRG approach aims to solve the nuclear many-body problem by performing a unitary transformation which maps the large-scale problem to effective operators acting on a tractable valence space. In implementing the unitary transformations, all intermediate operators are truncated at the two-body level, denoted IMSRG(2). This approximation is generally effective for energies, but misses highly collective quadrupole correlations [45]. The calculations were performed using two different sets of initial two-nucleon (NN) and three-nucleon (3N) forces derived from chiral effective field theory, indicated 1.8/2.0(EM) [46, 47] and N²LO_GO [48]. The former interaction is constrained by considering properties of two-, three-, and four-nucleon systems, and well reproduces energies throughout the medium and heavy region [49], but generally underpredicts charge radii. The latter one is additionally constrained by the saturation properties of nuclear matter which leads to a better reproduction of the absolute charge radii [49, 50].

MPMH and DFT calculations are based on effective in-medium interactions calibrated to nuclear observables across the nuclear chart. The MPMH calculations were performed using the Gogny D1M [51] effective interaction. The DFT Hartree-Fock-Bogoliubov (HFB) calculations were performed with two different energy density functionals: the Fayans functional Fy(IPV) [20] and the Skyrme functional SV-min [52]. For electromagnetic moments analysis, we also carried out the Hartree-Fock (HF) DFT calculations with the Skyrme functional UN-EDF1 [53] with the spin-spin term adjusted to reproduce the magnetic moment of $^{131}$In [19]. More details on the theoretical approaches used are provided in the Supplemental Material.

*Discussion.* All theoretical models provide similar wave function configurations for the high-spin states of $^{127,129}$In, based on a proton $g_{9/2}$ hole and a two-quasiparticle $d_{3/2}h_{11/2}$ neutron configuration, see Fig. 1(a) and Refs. [21, 27, 29]. As seen in Fig. 1, a reduction in the $\langle r^2\rangle$ values was observed for the high-spin isomer states relative to their ground states, resulting in a negative value of $\delta\langle r^2\rangle^{g,m}$. A variation of the charge radius relative to the g.s. radius has also been observed for other high-spin isomers across different regions of the nuclear chart [12, 13, 55–59]. Notably, the decrease observed for the high-spin isomer of $^{129}$In seen in Fig. 1(b) is about three times larger than the value observed for $^{127}$In, and comparable to the charge radius change between the g.s. configurations of $^{127}$In and $^{129}$In, $\delta\langle r^2\rangle^{127,129}$, shown in Fig. 1(d). This large reduction of the charge radius can only be explained by DFT calculations for the $I^\pi = 23/2^-$ maximally aligned state using the Fy(IVP) functional. The larger downshift of radii for the $I^\pi = 23/2^-$ state is suggested to be most likely caused by the additional gradient terms in the pairing and surface part of the Fayans functional. Such terms are absent in SV-min and HF models.

TABLE I. The measured $\delta\nu^{127,129}$ and $\delta\nu^{g,m}$ values and extracted $\delta\langle r^2\rangle^{127,129}$ and $\delta\langle r^2\rangle^{g,m}$ values. Experimental uncertainties are given in parentheses. Uncertainties from atomic theory calculations, shown in square brackets, were calculated following the approach presented in Ref [30]. The values for the isotope shifts and changes in the charge radii for both the ground-state $9/2$ and low-spin $1/2$ isomers can be found in Ref. [19]. For the high-spin state in $^{127}$In, values are given for two possible spin assignments of $(21/2^-)$ and $(23/2^-)$. $F$ and $K^{MS}$ values used to calculate $\delta\langle r^2\rangle^{A,A'}$ are taken from Ref. [20, 26].

| Transition | $F$ (GHz/fm$^2$) | $K^{MS}$ (GHz·u) | $\delta\nu^{127,129}$ (MHz) | $\delta\langle r^2\rangle^{127,129}$ (fm$^2$) | Mass | $I^\pi$ | $\delta\nu^{g,m}$ (MHz) | $\delta\langle r^2\rangle^{g,m}$ (fm$^2$) |
|---|---|---|---|---|---|---|---|---|
| 5p $^2$P$_{1/2}$ → 8s $^2$S$_{1/2}$ | 1.626[30] | 216[74] | 139(7) | 0.069(4)[6] | 127 | $(21/2^-)$ | $-29(7)$ | $-0.0179(43)[3]$ |
| | | | | | 127 | $(23/2^-)$ | $-21(7)$ | $-0.0129(43)[2]$ |
| | | | | | 129 | $23/2^-$ | $-87(7)$ | $-0.0535(43)[10]$ |
| 5p $^2$P$_{3/2}$ → 9s $^2$S$_{1/2}$ | 1.577[27] | 325[73] | 122(4) | 0.052(3)[6] | 127 | $(21/2^-)$ | $-19(5)$ | $-0.0121(32)[2]$ |
| | | | | | 127 | $(23/2^-)$ | $-20(6)$ | $-0.0128(36)[2]$ |
| | | | | | 129 | $23/2^-$ | $-80(3)$ | $-0.0508(18)[9]$ |
| **Average** | | | | **0.057(2)[6]** | 127 | $(21/2^-)$ | | $\mathbf{-0.0141(26)[3]}$ |
| | | | | | 127 | $(23/2^-)$ | | $\mathbf{-0.0128(28)[2]}$ |
| | | | | | 129 | $23/2^-$ | | $\mathbf{-0.0512(17)[9]}$ |

TABLE II. The hyperfine structure parameters $A_{hf}$ and $B_{hf}$ for the $^{127}$In and $^{129}$In isotopes, extracted magnetic moment $\mu$, spectroscopic quadrupole moment $Q_S$ and intrinsic quadrupole moment $Q_0$ values. The high-spin states were measured in this work. Statistical uncertainties (arising from experimental measurements) and systematic uncertainties (from atomic theory calculations) for $Q_S$ are given in parentheses and square brackets, respectively. For the high-spin state in $^{127}$In, values are given for two possible spin assignments of $(21/2^-)$ and $(23/2^-)$. The results for the ground states, $9/2^+$, were taken from Ref. [19].

| Mass | $I^\pi$ | $A_{hf}$ (MHz) | | | | $\mu$ ($\mu_N$) | $B_{hf}$ (MHz) | $Q_S$ (mb) | $Q_0^\dagger$ (mb) |
|---|---|---|---|---|---|---|---|---|---|
| | | 5p $^2$P$_{3/2}$ | 9s $^2$S$_{1/2}$ | 5p $^2$P$_{1/2}$ | 8s $^2$S$_{1/2}$ | | 5p $^2$P$_{3/2}$ | | |
| 127 | $9/2^+$ | 242(1) | 130(1) | 2278.3(6) | 243.8(4) | 5.532(2) | 338(16) | 587(28)[14] | 1076(53) |
| 129 | $9/2^+$ | 243(1) | 132(1) | 2304.9(9) | 244.8(7) | 5.596(2) | 280(7) | 487(13)[3] | 893(24) |
| 127 | $(21/2^-)$ | 99(1) | 54(1) | 942.5(7) | 100.9(7) | 5.340(4) | 464(12) | 805(21)[6] | 1058(29) |
| 127 | $(23/2^-)$ | 91(1) | 50(1) | 863.8(7) | 92.3(7) | 5.360(5) | 470(10) | 815(18)[6] | 1050(24) |
| 129 | $23/2^-$ | 101(1) | 55(1) | 956.6(6) | 101.3(4) | 5.935(4) | 344(18) | 598(32)[4] | 768(41) |

† The intrinsic quadrupole moments, $Q_0$, were estimated using $Q_0 = \frac{(I+1)(2I+3)}{I(2I-1)}Q_S$ [54].

As seen in Fig. 1(b) and Fig. 1(c), the DFT models predict similar, yet relatively large, differences in charge radii compared to the ground state for both high-spin isomers. In contrast, other models indicate significantly smaller changes in charge radii for both isomeric states. The DFT result for $I^\pi = 21/2^-$, however, has to be interpreted with caution because quasi-particle configuration of this state cannot be assigned unambiguously, see Supplemental Material. For all models, the predicted change in the g.s. charge radii $\delta\langle r^2\rangle^{127,129}$ is computed to be positive, consistent with experiment, but it exhibits a rather large model dependence.

Experimental results for the magnetic dipole and nuclear quadrupole moments are shown in Fig. 2. To separate the dependence on the nuclear spin, the intrinsic quadrupole moments, $Q_0 = \frac{(I+1)(2I+3)}{I(2I-1)}Q_S$, were extracted from the measured spectroscopic quadrupole moments $Q_S$, assuming a strong coupling scheme for these high-spin isomers [54]. They are shown in Fig. 2(c). In the case of HF and IMSRG, the intrinsic quadrupole moments were calculated independently, rather than extracted from the spectroscopic quadrupole moment. The extracted electromagnetic moments are compared with theoretical results in Fig. 2(a). All calculations were performed without explicit adjustment of effective charges or effective g-factors. All models reproduce the trend observed for the magnetic dipole moments, with larger deviations obtained for $^{127m}$In. For IMSRG, the magnetic moment calculations include the effect of two-body currents [60], which is essential to obtain a good agreement with the experiment. A comparison of the results obtained with and without two-body currents is shown in Fig 6 in the Supplemental Material. For HF, the agreement relies on adjusting the Landau parameter $g'_0$ in $^{131}$In [19].

In Fig 2(b) we compare our experimental (spectroscopic) quadrupole moments to the predicted values from DFT, MPMH, and IMSRG calculations. In general, the IMSRG and MPMH models largely underestimate the observed quadrupole moments. This is a well-known feature for the IMSRG calculations due to the IMSRG(2) truncation. For the MPMH model, the underestimation is due to the absent $\Delta N = 2$ quadrupole polarization which is neglected in the current valence space truncation scheme. The DFT calculations all reproduce the experimental moments rather well, especially for the $^{129}$In states and the $^{127}$In ground state. The quadrupole moments predicted by SV-min differ significantly from the values predicted by Fy(IVP) and HF calculations, with the appreciable increase of quadrupole moments in the high-spin states. This might be related to the pairing strength as stronger pairing reduces deformation effects. Indeed, SV-min predicts stronger pairing correlations than Fy(IVP), and HF has no pairing at all. In Fig 2(c) we compare the results obtained for the intrinsic quadrupole moments. The HF values reproduce the intrinsic moments for both $^{129}$In states very well. For $^{127}$In, a better agreement is observed if a spin $23/2^-$ is assumed for the isomeric state.

While some of the reduction of the charge radius of $^{129m}$In can be attributed to a reduced quadrupole collectivity in this excited state, this cannot account for the entire reduction. Indeed, a similar reduction in $Q_0$ is predicted by both Fy(IVP) and HF models, but their values of $\delta \langle r^2 \rangle^{g,m}$ are different. Moreover, the similar values of the charge radii of g.s. and isomeric state in $^{127}$In are not reproduced by Fy(IVP). However, as discussed in the Supplemental Material, the $21/2^-$ state cannot be unambiguously computed within a single-reference quasiparticle framework.

*Conclusions and outlook.* Our results reveal a significant reduction in the nuclear size and intrinsic quadrupole moments in the high-spin isomer $^{129}$In. The new data on isomeric electromagnetic moments were interpreted by complementary theoretical approaches that capture different aspects of nuclear structure. Except for Fy(IVP), our models cannot reproduce the magnitude of the charge radii reduction observed for the high-spin isomer of $^{129}$In. On the other hand, for the isomer $^{127m}$In, the Fy(IVP) model overestimates the observed value of the shift $\delta \langle r^2 \rangle^{g,m}$ for $^{127}$In, while DFT-HF and VS-IMSRG calculations predict a shift closer to the experimental value. The magnitude of the charge radius change between the ground states, $\delta \langle r^2 \rangle^{127,129}$, appears to be highly sensitive to the employed many-body methods, with DFT-Fy(IVP) and VS-IMSRG in closer agreement with experiment. The quadrupole moments are surprisingly well described by Fy(IVP) and HF approaches. Both VS-IMSRG and MPMH significantly underestimate the measured quadrupole moments. Based on our theoretical analysis, the reduction of $\delta \langle r^2 \rangle^{g,m}$ for $^{129}$In cannot be solely attributed to a shape effect.

IMSRG, MPMH, and HF calculations show good agreement with the measured nuclear magnetic dipole moments, suggesting a preference for an $I^\pi = 23/2^-$ assignment for $^{127m}$In. The quadrupole moments can be reproduced by Fy(IVP) and HF DFT calculations when using an unabridged single-particle space, allowing for the full development of shape polarisation. It is apparent that no single model can consistently describe the rich experimental data on electromagnetic moments in ground and isomeric states of $^{127,129}$In. The issue is expected to arise from the contrasting nature of the ground and high-spin states. A similar structure of the ground states in the neighboring isotopes will partially cancel systematic uncertainties in their charge radius differences, whereas such cancellation may not occur for their isomers. A consistent description will require accurate calculations for both ground and high-spin states and remains a theoretical challenge. Future measurements of high-spin isomers near doubly magic nuclei such as $^{56}$Ni, $^{100}$Sn, and $^{208}$Pb will be important for guiding further developments of nuclear theory and understanding the structure of nuclear isomers.

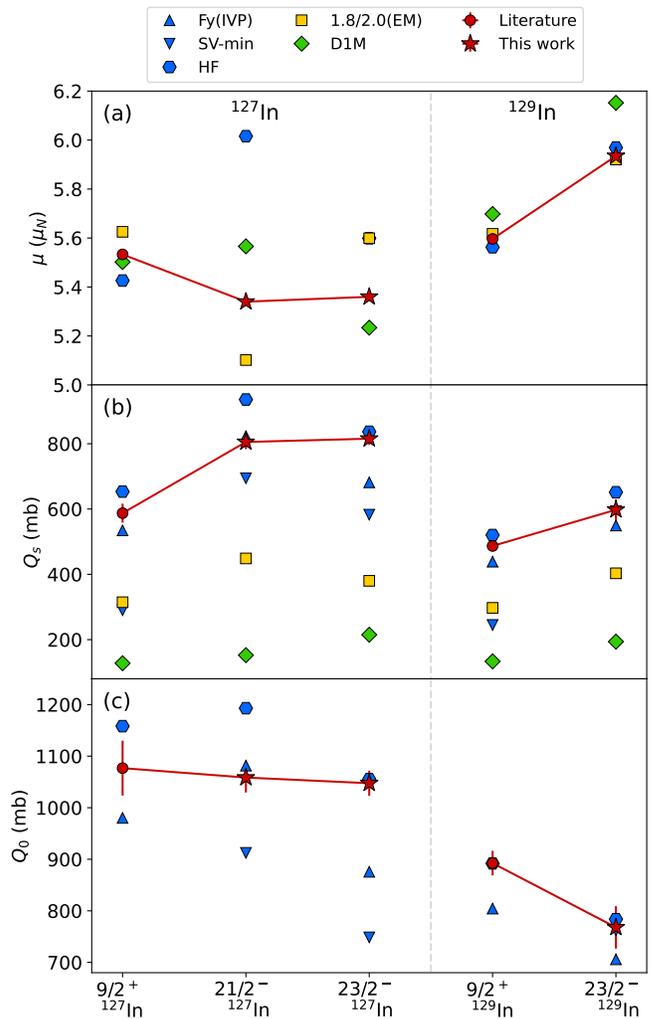

FIG. 2. Electromagnetic moments of the ground state and high-spin isomers of $^{127}$In and $^{129}$In: a) spectroscopic magnetic dipole moments, b) spectroscopic quadrupole moments, and c) intrinsic quadrupole moments. Experimental results (connected red markers) are compared with different calculations: DFT (blue markers), VS-IMSRG (gold markers) and MPMH (green markers). Results for the high-spin isomer of $^{127}$In are shown assuming two possible nuclear spin values of $21/2^-$ and $23/2^-$. Literature values for the ground states were taken from Ref. [19]. The line connecting the experimental results is for visual guidance only.

## ACKNOWLEDGEMENTS

This work was supported by ERC Consolidator Grant No. 648381 (FNPMLS); STFC grants ST/L005794/1, ST/L005786/1, ST/P004423/1, ST/P003885/1, ST/V001035/1, ST/V/001116/1 and Ernest Rutherford Grant No. ST/L002868/1; the U.S. Department of Energy, Office of Science, Office of Nuclear Physics under grants DE-SC0021176 and DE-SC0013365, and DE-SC0023175 (Office of Advanced Scientific Computing Research and Office of Nuclear Physics, Scientific Discovery through Advanced Computing); the BOF 14/22/104 from KU Leuven, BriX Research Program No. P7/12; the FWO-Vlaanderen (Belgium) project G080022N; the European Unions Grant Agreement 654002 (ENSAR2); National Key R&D Program of China (Contract No: 2018YFA0404403); the National Natural Science Foundation of China (No:11875073); a Leverhulme Trust Research Project Grant; the Polish National Science Centre under Contract No. 2018/31/B/ST2/02220; ERC under the European Union's Horizon 2020 research and innovation programme (Grant Agreement No. 101020842); the Deutsche Forschungsgemeinschaft (DFG, German Research Foundation) – Project-ID 279384907 – SFB 1245. We would also like to thank the ISOLDE technical group for their support and assistance, and the University of Jyväskylä for the use of the injection-locked cavity. We acknowledge the CSC-IT Center for Science Ltd., Finland, for the allocation of computational resources. This project was partly undertaken on the Viking Cluster, which is a high performance compute facility provided by the University of York. We are grateful for computational support from the University of York High Performance Computing service, Viking and the Research Computing team. B. K. S. acknowledges use of ParamVikram-1000 HPC facility at Physical Research Laboratory (PRL), Ahmedabad, for carrying out atomic calculations and his work at PRL supported by the Department of Space, Government of India. P.-G. R. thanks the regional computing center of the university Erlangen (RRZE) for support. The VS-IMSRG calculations were supported by NSERC under grants SAPIN-2018-00027 and RGPAS-2018-522453, the Arthur B. McDonald Canadian Astroparticle Physics Research Institute, and performed with an allocation of computing resources on Cedar at WestGrid and The Digital Research Alliance of Canada. K.M.L. acknowledges support from the Royal Society Dorothy Hodgkin Fellowship grant DHF\R1\231007 and UK Research and Innovation (UKRI) under the UK government's Horizon Europe funding Guarantee grant EP/Y036816/1 (ESPEN). The data that support the findings of this article are openly available [61–63].

\* Present affiliation: Duke University
† Email: rgarciar@mit.edu

## SUPPLEMENTAL MATERIAL

### Production of neutron-rich indium isotopes and high-spin isomers

The measured indium isotopes were produced at the CERN on-line isotope separator facility ISOLDE. Neutron-rich isotopes were produced by impingement of 1.4-GeV protons (accelerated by the CERN Proton Synchrotron Booster) onto the neutron converter (to suppress nearby caesium mass contamination [64]) of a uranium-carbide target [65]. The Cs contamination contributes to the non-resonant background observed in our spectra, thereby reducing our signal-to-noise ratio. The high energy and angular momentum of this process allow the population of high-spin isomer states to be produced, at a rate of just a few hundred ions/s for $^{127,129}$In. The indium isotopes diffused through the target material and were ionized by surface ionization, which was further enhanced by the resonant ionization laser ion source RILIS [66].

### The collinear resonance ionization setup

The produced indium ions were accelerated to 40 keV and mass separated using the ISOLDE high-resolution mass separator before being cooled and bunched using a gas-filled linear Paul trap (ISCOOL) [67, 68], with a bunch time of 10 ms. The ion bunches of 2 $\mu$s temporal width were then re-accelerated to 40034(1) eV and deflected into the CRIS beamline. The indium ions were then neutralised with a sodium-filled charge-exchange cell resulting in predicted relative atomic populations of 57% and 37% respectively for the 5p $^2P_{3/2}$ metastable state and 5p $^2P_{1/2}$ ground state [69]. The remaining ion fraction was removed by electrostatic deflectors, and the neutralized atom bunch was collinearly overlapped with two pulsed lasers, one for resonant excitation and another for non-resonant ionization. The atoms were excited using either the UV 246.0-nm transition (5p $^2P_{1/2} \to 8s\ ^2S_{1/2}$) or 246.8-nm transition (5p $^2P_{3/2} \to 9s\ ^2S_{3/2}$). The 5p $^2P_{1/2} \to 8s\ ^2S_{1/2}$ transition has increased sensitivity to magnetic dipole moments and isotope shifts, but is insensitive to $Q_S$, in contrast to the 5p $^2P_{3/2} \to 9s\ ^2S_{3/2}$ transition. The resonant laser light was produced by frequency tripling the light from an injection-locked Ti:Sapphire laser system [70]. This laser was seeded using a narrow-band SolsTiS continuous-wave Ti:Sapphire laser and pumped using a LEE LDP-100MQ Nd:YAG laser, producing pulsed narrow-band 738-nm laser light at 1 kHz. This light was then frequency tripled to 246.0 nm, producing 3 mW of laser light that was used to saturate the transition. The resonantly excited atoms were ionized by a final non-resonant 532-nm step, provided by a Litron LPY 601 50-100 PIV Nd:YAG laser at 100 Hz. The frequency of the resonant first step was scanned and the resulting ions were deflected onto a detector, producing the hyperfine spectra from which the hyperfine parameters and isomer shifts were obtained. Typical spectra are shown in Figs. 3 and 4. The wavelengths were measured using a HighFinesse WSU2 wavemeter, which was drift stabilized by simultaneous measurement of a Toptica DLC DL PRO 780 diode laser locked to the 5s$^2S_{1/2} \to 5p\ ^2P_{3/2}$ F = 2 - 3 transition of $^{87}$Rb using a saturated absorption spectroscopy unit [72].

### Example hyperfine spectra

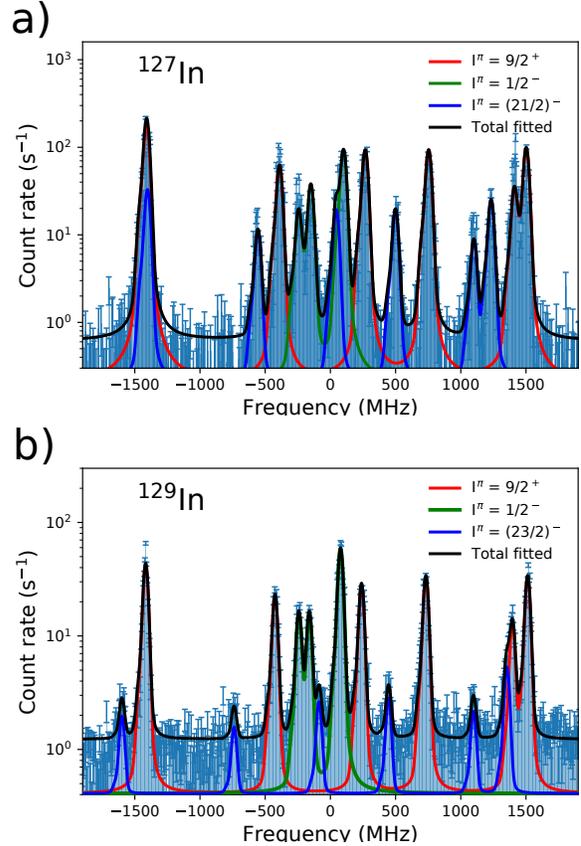

FIG. 3. Hyperfine spectra obtained using the 5p $^2P_{3/2} \to 9s\ ^2S_{1/2}$ transition for (a) $^{127}$In and (b) $^{129}$In.

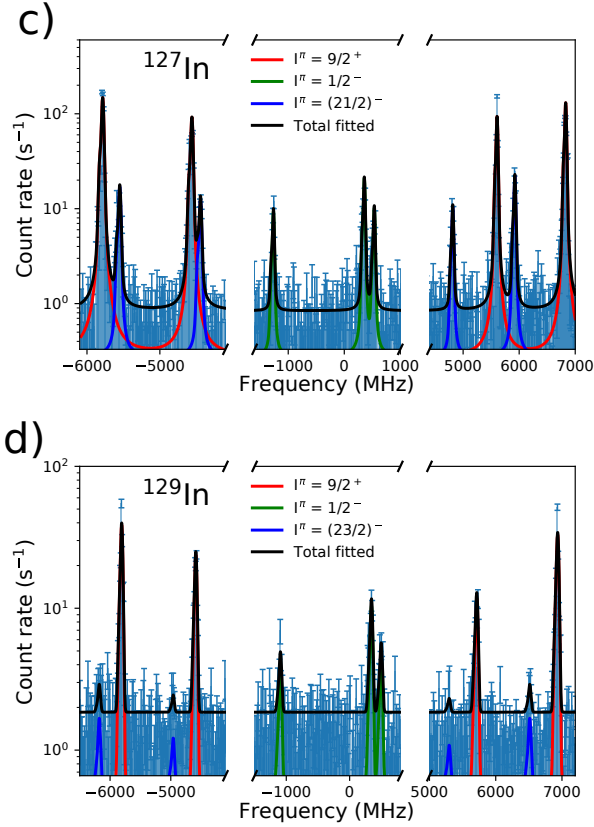

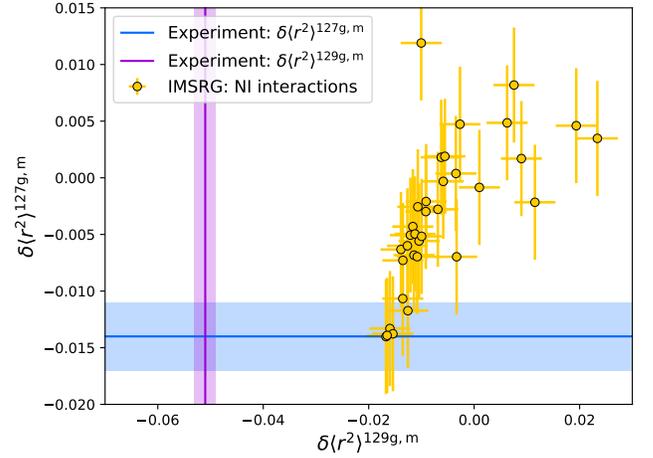

FIG. 5. Theoretical change in charge radii values (markers) for the high-spin state, relative to the ground state, for $^{127}$In and $^{129}$In calculated with the VS-IMSRG method using 34 non-implausible interactions [77]. These are compared to experimental $\delta\langle r^2\rangle^{g,m}$ values (lines) for $^{127}$In (blue) and $^{129}$In (purple).

FIG. 4. Hyperfine spectra obtained using the $5p\ ^2P_{1/2} \rightarrow 8s\ ^2S_{1/2}$ transition for (c) $^{127}$In and (d) $^{129}$In.

## VS-IMSRG calculations

The VS-IMSRG calculations were performed, starting from the underlying NN and 3N interactions expressed in the harmonic oscillator basis. This was used to construct the effective Hamiltonian within a valence space where the exact diagonalization is feasible [37]. The effective Hamiltonian was decoupled through an approximately unitary transformation derived from the Magnus expansion method [73]. Applying the same unitary transformation to operators for electromagnetic moments and radii yields effective valence-space operators consistent with the Hamiltonian. During the IMSRG transformation, all the operators which arise are truncated at the two-body level. The decoupling was performed with the multi-shell approach [74] starting from a 15 major shell harmonic oscillator space with frequency $\hbar\omega = 16$ MeV, taking the proton $\{1p_{1/2},\ 1p_{3/2},\ 0f_{5/2},\ 0g_{9/2}\}$ and neutron $\{2s_{1/2},\ 1d_{3/2},\ 1d_{5/2},\ 0g_{7/2},\ 0h_{11/2}\}$ valence space orbitals above the $^{78}$Ni core. Note that in a sufficiently large single-particle space, results are independent of $\hbar\omega$.

For the initial three-body force, the additional truncation was introduced using the sum of the three-body harmonic oscillator quanta, and with a recently introduced novel storage scheme [75], allows us to use 22 in this work, more than sufficient to achieve converged results with respect to this parameter. The valence space diagonalization and evaluation of charge radii were done with the KSHELL code [76].

To investigate why the isomer shift of $^{129}$In might be significantly underestimated, we performed the VS-IMSRG calculations using 34 different interactions that were presented in the context of predicting the neutron skin of $^{208}$Pb [77]. The results are shown in Fig. 5. The uncertainties include the convergence of the model space and the truncation of effective field theory (EFT). The model-space error is estimated from the difference between the 13 and 15 major-shell calculations. The EFT truncation error is assigned as the difference between the results with the NLO and NNLO delta-full interactions [78]. As illustrated in the figure, none of the considered interactions are capable of accurately predicting the isomer shifts for both $^{127}$In and $^{129}$In at the same time, reinforcing the suspicion that the disagreement is due to the IMSRG(2) truncation.

Fig. 6(a) presents calculations of the magnetic moments obtained using only one-body currents, and those where two-body currents are included. As illustrated in the figure, the inclusion of two-body currents, which were recently implemented in VS-IMSRG calculations [60], is essential to achieve good agreement with experimental results. Fig. 6(b) presents calculations of the intrinsic quadrupole moments. The quadrupole sum rule [79] is employed to compute the intrinsic quadrupole moment (markers) and its variance (shaded bands), corresponding to the intrinsic deformation and its rigidness, respectively. A smaller variance corresponds to a steeper potential energy surface.

## MPMH calculations

The MPMH calculations are based on a Multi-Configuration Self-Consistent field approach which determines a priori in a self-consistent way the N-body wave functions of CI type and the underlying one-body states, thus preserving the variational character. While the orbitals equation is solved for the whole set of orbitals that are expanded on a harmonic oscillator (HO) basis of size $N_{shell}$, the CI configurations are built using a given truncation scheme. The MPMH calculations were performed with D1M Gogny parameterization [51]. The calculations carried out with the D1S Gogny force [80, 81] yield very similar results.

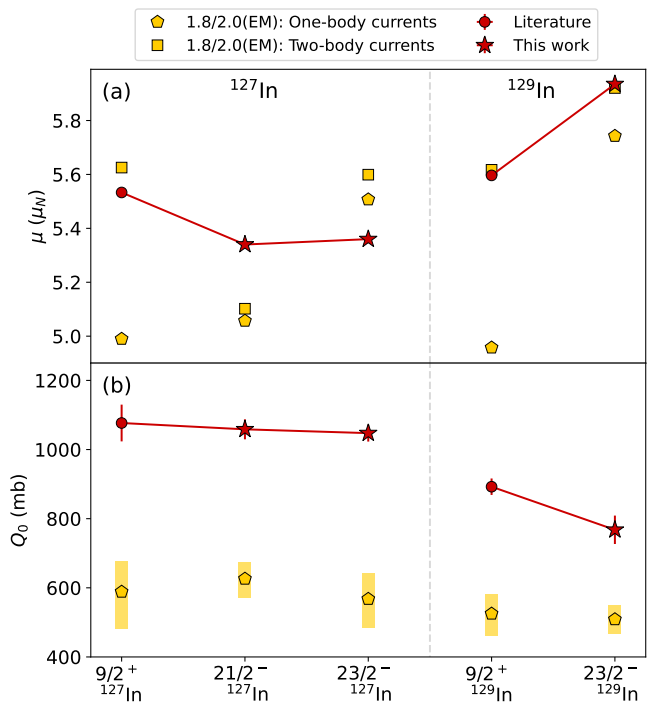

FIG. 6. (a) Magnetic moments and (b) intrinsic quadrupole moments of the ground state and high-spin isomers for $^{127}$In and $^{129}$In. Experimental results (connected red markers) are compared with VS-IMSRG calculations (gold markers) including one-body (pentagons) and two-body currents (squares). Results for the high-spin isomer of $^{127}$In are shown assuming two possible nuclear spin values of $21/2^-$ and $23/2^-$. Literature values for the ground states were taken from Ref. [19]. The red line connecting the experimental results is for visual guidance only and does not indicate a trend. Shaded bands for the intrinsic quadrupole moment relate to the rigidness of the deformation. See text for details.

In the present calculations, a HO basis of $N_{shell} = 11$ size has been used. To build the N-body wave functions, the model space included the proton $\{1p_{1/2}, 0g_{9/2}, 0g_{7/2}, 1d_{5/2}, 1d_{3/2}, 2s_{1/2}\}$ and neutron $\{0g_{9/2}, 0g_{7/2}, 1d_{5/2}, 1d_{3/2}, 2s_{1/2}, 0h_{11/2}\}$ orbitals. Indeed, simple constrained Hartree-Fock-Bogoliubov (HFB) calculations with $N_{shell} = 11$ revealed that both $^{127}$In and $^{129}$In ground states display an axial softness around sphericity with a small intrinsic prolate deformation. Thus, valence spaces were chosen in order to include spherical orbitals which are linked to the axially deformed ones in the minimum energy area and close to the Fermi levels. Moreover, in the proton sector, configurations up to 3p3h were taken into account. In the neutron sector, configurations up to 2p2h and 4p4h were included for $^{129}$In and $^{127}$In, respectively. Finally, in the present calculations, the single-particle orbitals used were the Hartree-Fock calculations and no effective charges were used.

In this study, MPMH gives satisfactory results for the isomer excitation energies and M1 properties as those observables depend essentially on $\Delta N = 0$ configurations. The present MPMH calculations suggest a reduction of collectivity and nuclear size for the high-spin isomer, however, the magnitude of the charge radius reduction is not reproduced. This is due, in particular, to the absence of $\Delta N = 2$ configurations. One may invoke also a lack of spin-orbit effect that may come directly from the spin-orbit term itself or a tensor term which is not included in the D1M parameterization. Moreover, the renormalization of the orbitals may play a non negligible role in the nuclear size reduction. In the MPMH results, one observes the decrease of the correlation energies $E_{corr}$ in the isomers (defined as the difference between the spherical HF and MPMH binding energies) that is caused by the blocking mechanism which hinders neutron pairing: $E_{corr}^{g.s.}(^{129}In) \simeq 1.3$ MeV, $E_{corr}^m(^{129}In) \simeq 0.7$ MeV, $E_{corr}^{g.s.}(^{127}In) \simeq 2.0$ MeV and $E_{corr}^{m.}(^{127}In) \simeq 1.6$ MeV. The global larger correlation energies in $^{127}$In come from stronger neutron pairing-type correlations. Whatever the isotope, the state and the parameterization of the Gogny interaction, proton occupation probabilities are found very similar. Only $\simeq 0.1$ proton is scattered beyond the $0g_{9/2}$-$0g_{7/2}$ gap. The neutron $0h_{11/2}$ occupation probability is increased by $\simeq 0.1$ in the correlated ground states compared to the HF calculation. The structure of the isomers is explained by the migration of $\simeq 1.0$ neutron from the $1d_{3/2}$ to the $0h_{11/2}$ orbital. The decrease of the correlation energy in the isomers is caused by this blocking mechanism which hinders neutron pairing. The same effect was found in HFB calculations for which ground states are calculated as one blocked proton quasi-particle state and isomers as one blocked proton and two blocked neutron quasiparticles. In that case, one noted the absence of proton pairing correlation in all the states and a strong reduction of neutron pairing energy between the ground state and the isomer (from $\simeq 10$ MeV to $\simeq 2$ MeV in $^{127}$In and from $\simeq 6$ MeV to zero in $^{129}$In).

## DFT calculations

In this work, we used three different variants of the nuclear DFT approach. Results were obtained for the SV-min functional [52] and BCS pairing with the solver `SkyAx` for axially symmetric deformed configurations [82]. The same solver was used for the Fayans Fy(IVP) functional [20] and HFB pairing. For the UNEDF1 functional [53], we carried out no-pairing HF calculations using version (v3.12e) of solver `HFODD` [83, 84]. All DFT variants allow time-reversal symmetry breaking in the computation of broken-pair states.

The Fayans functionals differ from Skyrme functionals mainly by having a gradient term in pairing and in the surface energy. This allowed, for the first time, to tune radius differences in isotopic chains [85, 86] using the same pairing strengths for protons and neutrons. The recently calibrated Fy(IVP) functional is more flexible in that it allows for separate proton and neutron pairing strengths, which improves the performance of radius differences for heavier isotopic chains e.g. Sn and Pb [20].

In Fy(IVP) and SV-min DFT calculations, the charge radii were obtained directly from the charge form factors including relativistic corrections (including spin-orbit term) and contributions from nucleonic charge form factors [87]. Since in solver `HFODD` those corrections were not yet implemented, in the UNEDF1 calculations the HF proton radii were used instead. For the differences of radii analyzed in this work, we estimated the effect of such an omission to be of the order of $\pm 0.007$ fm.

For the high-spin isomers discussed in this work, three quasi-particle states were selected, in which one proton pair and two neutron pairs were broken, see Fig. 1 and Table III. For the pairing calculations, those states were calculated using the standard blocking method [88], with the blocked states corresponding to the creation of single-particle hole states in the paired vacua. In the HF calculations, the same broken-pair hole states were created in the closed proton $Z = 50$ and neutron $N = 82$ shells.

In all variants of the DFT calculations, the selected configurations were determined self-consistently, whereupon the full shape (deformation) and spin-polarisation effects were automatically taken into account. In addition, the angular-momentum symmetry was restored [89] in the UNEDF1 HF calculations, which allowed for the determination of the spectroscopic magnetic dipole and spectroscopic electric quadrupole moments that are shown in Fig. 2. This was particularly important for the spectroscopic magnetic dipole moments, which cannot be inferred from the intrinsic moments [44].

Table III shows details of the DFT configurations used in the present work. Rows of Table III list the Kramers-degenerate spherical single-particle orbitals described by the (a) spherical $j, K$ or (b) dominating Nilsson $[Nn_z\Lambda]K$ quantum numbers. Quantum numbers $K = |j_z|$ represent moduli of the angular-momentum projections on the $z$-axis. As it turns out, even for the spherical single-particle states, the asymptotic Nilsson quantum numbers, which become exact only at large deformations, constitute useful indicators that allow us to label those states uniquely. For each configuration listed in columns (c)-(i), numerical values indicate projections $\Omega = \pm K$ of the blocked single-particle states. Entries pair(0) or pair(2) denote orbitals with both projections $\pm K$ paired and occupied symmetrically. For the unpaired HF calculations, they correspond to 0 or 2 particles occupying the given orbital, respectively. The last row of Table III shows the total angular momentum and parity of each configuration.

It is important to note that for every self-consistently converged configuration, the spin and orbital parts of the blocked-state angular-momentum projections depart from the spherical values and spread among all other paired orbitals of the core. In particular, blocked neutron states can generate proton core contributions and vice versa. This polarisation mechanism is responsible for the spin core polarisation and generates the essential core contributions to magnetic moments. It relies on the fact that for the broken time-reversal, the paired states (the canonical pairs [88]) are not built of time-reversed single-particle states and thus can carry non-zero angular momentum.

Columns (c)-(g) list the five simplest ground-state and isomeric configurations discussed in the text, with two possible (unmeasured) angular momenta that can be assigned to the $^{127m}$In isomer, $21/2^-$ and $23/2^-$. In addition, columns (h)-(i) list two other possible $21/2^-$ configurations of that isomer, indicating that a configuration-mixing calculation (not performed in this study) could be required.

TABLE III. Configurations of ground and isomeric states in $^{127}$In and $^{129}$In considered in this work (see text for details).

| (a) Orbital | (b) $[Nn_z\Lambda]K$ | (c) $^{127}$In | (d) $^{127m}$In | (e) $^{127m}$In | (f) $^{129}$In | (g) $^{129m}$In | (h) $^{127m}$In | (i) $^{127m}$In |
|---|---|---|---|---|---|---|---|---|
| $\pi 1\mathrm{g}_{9/2,9/2}$ | $[404]9/2$ | $+9/2$ | $+9/2$ | $+9/2$ | $+9/2$ | $+9/2$ | $+9/2$ | $+9/2$ |
| $\nu 1\mathrm{h}_{11/2,11/2}$ | $[505]11/2$ | pair(0) | pair(0) | $+11/2$ | pair(2) | $+11/2$ | $+11/2$ | $+11/2$ |
| $\nu 1\mathrm{h}_{11/2,9/2}$ | $[514]9/2$ | pair(2) | $+9/2$ | pair(0) | pair(2) | pair(2) | pair(0) | pair(0) |
| $\nu 2\mathrm{d}_{3/2,3/2}$ | $[402]3/2$ | pair(2) | $+3/2$ | $+3/2$ | pair(2) | $+3/2$ | pair(2) | pair(2) |
| $\nu 2\mathrm{d}_{3/2,1/2}$ | $[400]1/2$ | pair(2) | pair(2) | pair(2) | pair(2) | pair(2) | $+1/2$ | pair(2) |
| $\nu 3\mathrm{s}_{1/2,1/2}$ | $[411]1/2$ | pair(2) | pair(2) | pair(2) | pair(2) | pair(2) | pair(2) | $+1/2$ |
| $I^\pi$ | | $9/2^+$ | $21/2^-$ | $23/2^-$ | $9/2^+$ | $23/2^-$ | $21/2^-$ | $21/2^-$ |